\documentclass{aastex6}
\usepackage{natbib}
\usepackage{gensymb}
\usepackage{amsmath}
\usepackage{url}
\usepackage{multirow}
\usepackage{bm}
\newcommand{\myemail}{lilin@shao.ac.cn}

\shorttitle{The Galactic extinction and reddening from SCUSS}
\shortauthors{Li et al.}
\begin{document}
\title{The Galactic extinction and reddening from the South Galactic Cap  \textnormal{$u$}-band Sky Survey: u band galaxy number counts and $u-r$ color distribution}

\author{Linlin Li \altaffilmark{1,2}, Shiyin Shen \altaffilmark{1,3}\dag, Jinliang Hou\altaffilmark{1}, Fangting Yuan \altaffilmark{1}, Jing Zhong \altaffilmark{1}, Hu Zou \altaffilmark{4}, Xu Zhou \altaffilmark{4}, Zhaoji Jiang \altaffilmark{4}, Xiyan Peng \altaffilmark{4}, Dongwei Fan \altaffilmark{4}, Xiaohui Fan \altaffilmark{4}, Zhou Fan \altaffilmark{4}, Boliang He\altaffilmark{4}, Yipeng Jing \altaffilmark{6}, Michael Lesser \altaffilmark{5}, Cheng Li \altaffilmark{1}, Jun Ma \altaffilmark{4}, Jundan Nie \altaffilmark{4}, Jiali Wang \altaffilmark{4}, Zhenyu Wu \altaffilmark{4}, Tianmeng Zhang \altaffilmark{4}, Zhimin Zhou \altaffilmark{4}}

\altaffiltext{1}{Key Laboratory for Research in Galaxies and Cosmology, Shanghai Astronomical Observatory, Chinese Academy of Sciences, 80 Nandan \\Road, Shanghai, China, 20030; \myemail}
\altaffiltext{2}{University of Chinese Academy of Sciences, 19A Yuquanlu, Beijing, China, 100049}
\altaffiltext{3} {Key Lab for Astrophysics, Shanghai 200234 China; ssy@shao.ac.cn}
\altaffiltext{4}{Key Laboratory of Optical Astronomy, National Astronomical Observatories, Chinese Academy of Sciences, Beijing, 100012, China;}
\altaffiltext{5}{Steward Observatory, University of Arizona, Tucson, AZ 85721, USA}
\altaffiltext{6}{Center for Astronomy and Astrophysics, Department of Physics and Astronomy, Shanghai Jiao Tong University, Shanghai 200240, China}
\altaffiltext{\dag}{ssy@shao.ac.cn}

\begin{abstract}

We study the integral Galactic extinction and reddening based on the galaxy catalog of the South Galactic Cap U-band Sky Survey (SCUSS), where $u$ band galaxy number counts and $u-r$ color distribution are used to derive the Galactic extinction and reddening respectively. We compare these independent statistical measurements with the reddening map of \citet{Schlegel1998}(SFD) and find that  both the extinction and reddening  from the number counts and color distribution are in good agreement with the SFD results at low extinction regions ($E(B-V)^{SFD}<0.12$ mag). However, for high extinction regions ($E(B-V)^{SFD}>0.12$ mag), the  SFD map overestimates the Galactic reddening systematically, which can be approximated  by a linear relation $\Delta E(B-V)= 0.43[E(B-V)^{SFD}-0.12$]. By combing the results of galaxy number counts and color distribution together, we find that the shape of the Galactic extinction curve is in good agreement with the standard $R_V=3.1$ extinction law of \cite{ODonnell1994}.

\end{abstract}


\section{Introduction}
Interstellar dust absorbs and scatters photons from most astronomical sources \citep{Draine2003}, which causes extinction and reddening in observations. All the extragalactic objects are extincted and reddened by the Milky Way interstellar dust \citep{Oguri2014,Denney2010}. A map of the integral Galactic dust extinction is essential for all  extragalactic studies.

The  most commonly used all-sky dust extinction map was constructed by \cite{Schlegel1998} (hereafter SFD), which was derived from the composite of the COBE/DIRBE and IRAS/ISSA infrared maps and the Leiden-Dwingeloo map of HI emission. With $E(B-V)$ values provided by the SFD map, the Galactic extinction in any given photometric band can be parameterized  by $A_{\lambda}=k(\lambda)E(B-V)$, where $k(\lambda)$ is the extinction coefficient determined by the Galactic extinction curve (see \citealt{Cardelli1989} and it's update \citealt{ODonnell1994}, hereafter ODO).

The $E(B-V)$ values of the SFD map have been tested  by many independent measurements. \cite{Arce1999} used four different methods to derive the extinction in the Taurus dark cloud complex, suggesting that the SFD map overestimates the extinction by a factor of $1.3-1.4$ in regions where $A_{\rm V}>0.5$ mag \citep[see also][]{Dobashi2005,Schlafly2010,Schlafly2011,Yuan2013}. However, these tests are typically based on spectral types of Galactic stars. The integral line-of-sight dust to stars is not directly comparable to, but the lower limit of the integral Galactic extinction of extragalactic sources.

Galaxy number count is an independent approach that can  test the integral Galactic extinction. Under the assumption of an isotropic galaxy distribution on large scales, there will be smaller number of galaxies in the direction of higher Galactic extinction down to the same apparent magnitude \citep{Burstein1982, Fukugita2004,Yasuda2007}. Using the $r$ band galaxy number counts in the  Sloan Digital Sky Survey \citep[SDSS;][]{York2000}, \cite{Yasuda2007} found that  the SFD map overestimates the Galactic extinction in high extinction regions ($E(B-V)>0.15$ mag).

Besides extinction effect, Galactic dust also reddens the colors of background galaxies \citep{Gonzlez1999,Schrder2007,Peek2010}. Compared with the number counts of galaxies, the average color relies less on the assumption of a homogeneous distribution of galaxies. Using passive galaxies as color standards, \citet{Peek2010} estimated the Galactic reddening and, on the contrary, found that the SFD map under-predicts the reddening toward low Galactic latitudes (high extinction regions). However, as claimed in \citet{Peek2010}, this result does not necessarily conflict with the galaxy number counts ones since the regions in this study all have $E(B-V)<0.15$.

To have a better constraint on the Galactic extinction and resolve  possible conflicts between the studies of galaxy number counts and galaxy colors, a high quality photometry data in shorter wave-length (e.g. $u$ band) is helpful where the extinction/reddening effect is more significant. For the SDSS data, the photometry depth and accuracy in $u$ band is not comparable to the other bands due to its low efficiency. Alternatively, the South Galactic Cap $u$-band Sky Survey (SCUSS) \citep{Zhou2015}, which is a $u$ band (354nm) survey in high latitude region of the South Galactic region, and provides $u$ band data with a depth of about 1.5 magnitude deeper than the SDSS one. In this study, we take the released catalog of SCUSS \citep{Zou2016} to study the Galactic extinction with both the $u$ band galaxy number counts and $u-r$ color distribution by further combining the SDSS data. Our motivation is to take the advantages of the SCUSS $u$ band data and  make a detailed and self-consist statistical study on the Galactic extinction and reddening in the south Galactic cap region.

This paper is organized as follows. In Section \ref{sec2}, we introduce the galaxy catalog related to this study. In Section \ref{sec3} and Section \ref{sec4} , we study the Galactic extinction and reddening using the $u$ band galaxy number counts and $u-r$ color distribution respectively. We discuss the Galactic extinction curve in Section \ref{sec5}. Finally, we give a brief summary in Section \ref{sec6}.

\section{Data} \label{sec2}

\subsection{SCUSS photometry}
SCUSS is a deep (deeper than SDSS) $u$ band image survey in the north part of the south Galactic cap region. The effective wavelength of the SCUSS $u$ band filter is 3538 \AA , slightly different from (bluer than) the SDSS $u$ band. The survey is undertaken on the 2.3m Bok telescope at Kitt Peak with a 4k$\times$4k CCD camera, which provides a field of view 1\arcdeg.08$\times$1\arcdeg.03 and  image resolution of 0.454 arcsec per pixel. The footprint of the survey is initially designed to be the region where $b < -30\arcdeg$ and $\delta> -10\arcdeg$ and later slightly extended to lower Galactic latitudes. Each field typically has two exposures and the total exposure time amounts to more than 5 minutes, which provides  photometry depth of about 1.5 mag deeper than the SDSS $u$ band data (see more discussions in Section 2.3).
The overview of the survey can be found in \cite{Zhou2015} and visited from \url{http://batc.bao.ac.cn/Uband/}. In this study, we take the released data of SCUSS, which includes 3700 fields and with a total sky coverage of about 4115 deg$^2$ \citep{Zou2016}.
The details of the data reduction of SCUSS images can be found in \cite{Zou2015}. Here, we list some key ingredients of the data reduction related to this study.

The single-epoch images are first stacked to form a combined image for each field. SExtractor \citep{Bertin1996} is then used to detect objects on the stacked images. The detected sources are named as ``total sources'' with  ``automatic magnitude'' recorded, which are also classified as point/extended sources according to ``BERTIN\_G\_S''.
For these fields also with SDSS photometry, the ``total objects'' were then matched with the objects in the SDSS Data Release 9 (DR9) catalog with any of the $u,g,r,i,z$ magnitudes measured. All the matched objects are noted as ``core sources''. For these ``core sources',  the flag ``Type'' in SDSS $r$ band replaces the ``BERTIN\_G\_S'' and  is used to make the star/galaxy classification. For galaxies in ``core sources'', their SDSS $r$ band de Vaucouleurs and exponential likelihoods are further adopted to derive the model magnitude (``modelMag'') in SCUSS $u$-band as for the ``modelMag'' in other SDSS bands.

In this study, we take ``total sources''  for $u$ band galaxy number counts (Section \ref{sec3}) and use  ``core sources'' for reddening measurements from $u-r$ colors (Section \ref{sec4}).
Since we aim to study the Galactic extinction using the extragalactic sources, all the magnitudes of galaxies have not been corrected for Galactic extinction.

\subsection{Galaxy sample}
In this study, we take the galaxies from the region where SCUSS overlaps with SDSS. The SDSS footprint in the south Galactic region is about 5192 deg$^{2}$ and its photometry catalog had been released in DR8 \citep{Aihara2011}. Among 3700 SCUSS fields, 3070 are fully covered by SDSS and the total sky coverage is 3415 deg$^{2}$.

To ensure the uniformity of the photometry of the final sample of galaxies, we make further detailed selections on both  the SCUSS and SDSS data.

\begin{description}
\item[Bright star mask]\hfill \\

Bright stars contaminate the photometry of both  the SCUSS and SDSS data. We use the bright star masks of the BOSS tilling geometry \footnote{\url{http://www.sdss3.org/dr9/algorithms/boss_tiling.php}} to remove the area contaminated by the bright stars \citep{Blanton2003}. For 3415 deg$^{2}$ \textbf{SCUSS/SDSS} area, the total area inside the bright star masks amounts to 33.88 deg$^{2}$.

\item[SCUSS exposure time selection]\hfill \\

In each field of the SCUSS, even for the combined image, not all the areas have been covered by 2 exposures because of the gaps in the CCD camera \citep{Zou2015}. Figure~\ref{maskimage} shows an example of the exposure map for a typical SCUSS field. As can be seen, there are few stripes at the CCD gap regions with only one exposure, whereas few overlapped regions have more than two exposures. To ensure the uniform optical depth in the galaxy number counts, we take the survey exposure map and only select galaxies inside the field with more than one exposure. With such a selection, the sky coverage of SCUSS/SDSS data is further reduced to 2987 deg$^{2}$.
\begin{figure}
\epsscale{1}
\includegraphics[scale=0.6]{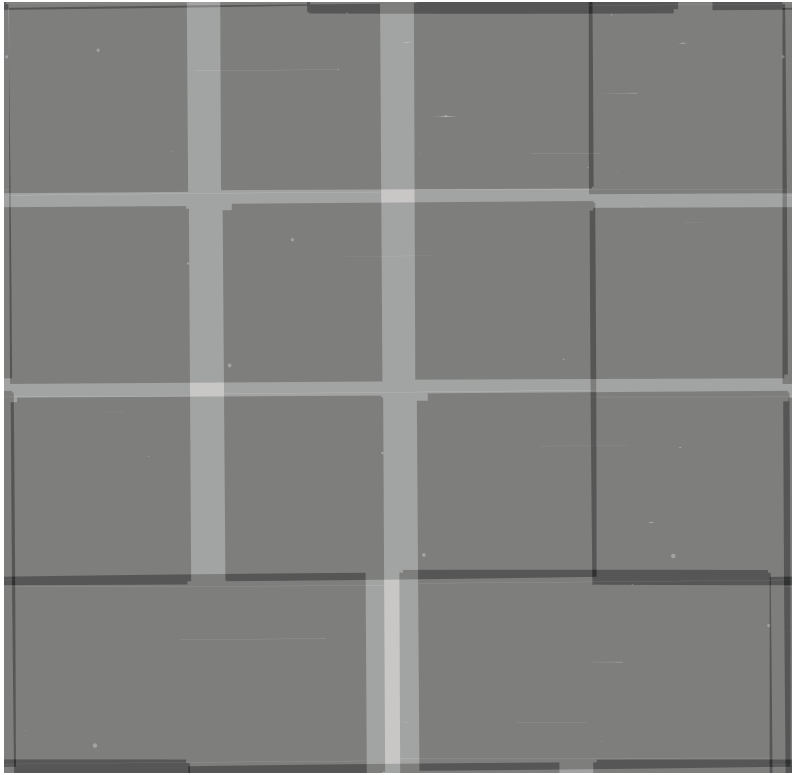}
\centering
\caption{The exposure map of an example SCUSS field. The dark gray area (most occupied) has two exposures, while the white, white gray and black areas have 0, 1 and $>2$ exposures respectively.
\label{maskimage}}
\end{figure}

\item[SDSS poor photometry]  \hfill \\

In SDSS, there are image masks indicating poor photometry \footnote{\url{http://data.sdss3.org/sas/dr9/boss/lss/badfield_mask_unphot-ugriz_pix.ply}}. Inside the SDSS/SCUSS overlapping area (2987 deg$^2$ after removing the SCUSS gaps), the total area with poor photometry mask adds up to 18.17 deg$^2$. For SCUSS, the photometry of all fields are good since all poor fields have been observed repeatedly.

\end{description}

In summary, when we do the galaxy number counts in SCUSS, we only select galaxies in the area with at lease 2 exposures and avoid the area contaminated by bright stars. When we include the SDSS photometry to study the $u-r$  colors in Section \ref{sec4}, we further exclude the galaxies within the SDSS poor photometry mask.  We take ``automatic magnitude'' for galaxy number counts  and use  ``model magnitude'' when studying their $u-r$ colors.

\subsection{Galaxy number counts in SCUSS}
\begin{figure}
\epsscale{1}
\includegraphics[scale=0.6]{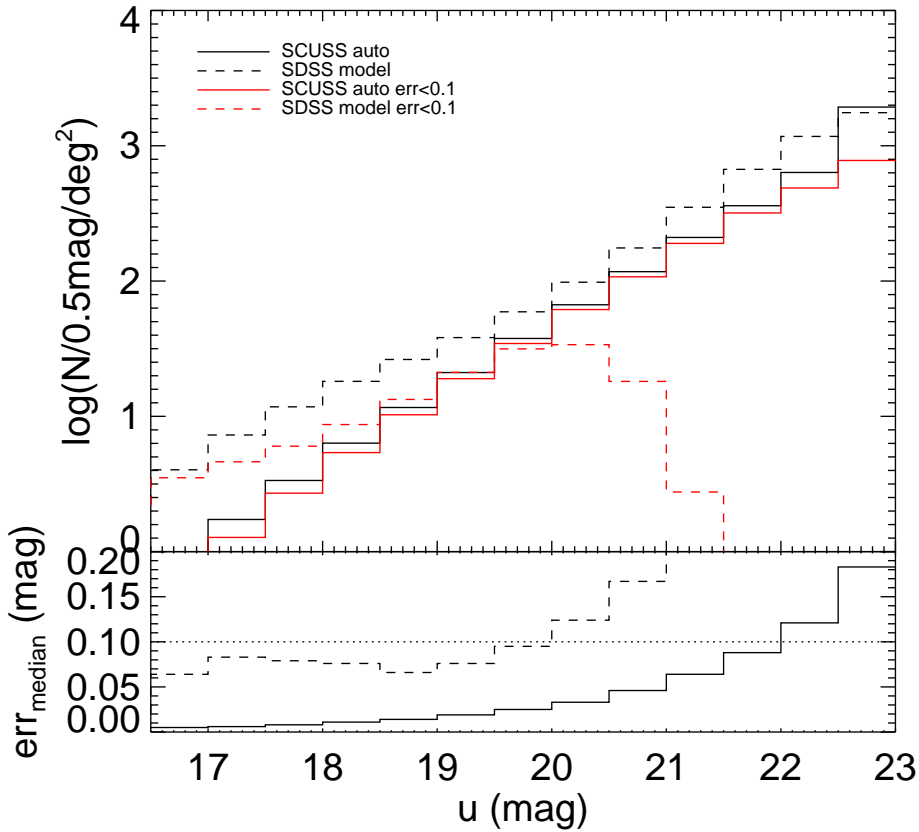}
\centering
\caption{Number counts and photometric errors of galaxies in SCUSS and SDSS. Top panel: the black solid and dashed histograms show the number counts of galaxies in SCUSS and SDSS respectively, while the red solid and dashed histograms further show the galaxies with photometric error $<0.1$. Bottom panel: the median of photometric errors of SCUSS (solid) and SDSS (dashed) galaxies. A horizontal line $err_{median}=0.1$ is shown for illustration.
\label{histerr}}
\end{figure}

In Figure~\ref{histerr}, the top panel shows the histogram of the galaxy number counts in SCUSS $u$ band whereas the bottom panel shows the median photometric error in each given magnitude bin (solid histograms). We also plot the SDSS $u$ band results as the dashed histograms for comparison.

Apparently, the number counts of galaxies in SDSS $u$ band even outnumbers the SCUSS data (up panel). The main reason of this outnumbering is that the ``modelMag'' in $u$ band of SDSS is a forced  measurement based on the profile model of SDSS $r$ band which is even deeper than the SCUSS $u$ band data for typical galaxies (see $u-r$ color distribution of galaxies in Section \ref{sec4}). Because of this, the SDSS $u$ band data have significantly larger photometric errors than the SCUSS sources (bottom panel). To further illustrate this, we also show the number counts of the SDSS and SCUSS $u$ band galaxies with photometric error smaller than 0.1 mag as the red dashed and red solid histograms in the top panel. After removing the objects with high photometric uncertainty, the SDSS $u$ band number counts match the SCUSS data at bright end ($u<20$) more closely. However, the galaxies in SDSS $u$ band still outnumbers the SCUSS catalog at very bright end ($r<18$ mag). To further verify these excess detections in SDSS, we have made visual inspections on both of the SDSS and SCUSS $u$ band images. We find that, except  fake detections from  diffraction spikes of bright stars, most of these `bright objects' are caused by the forced fitting of the model magnitude with unrealistic large apertures.

\section{Galactic extinction in $u$ band} \label{sec3}
\begin{figure}
\figurenum{3}
\centering
\includegraphics[scale=0.4]{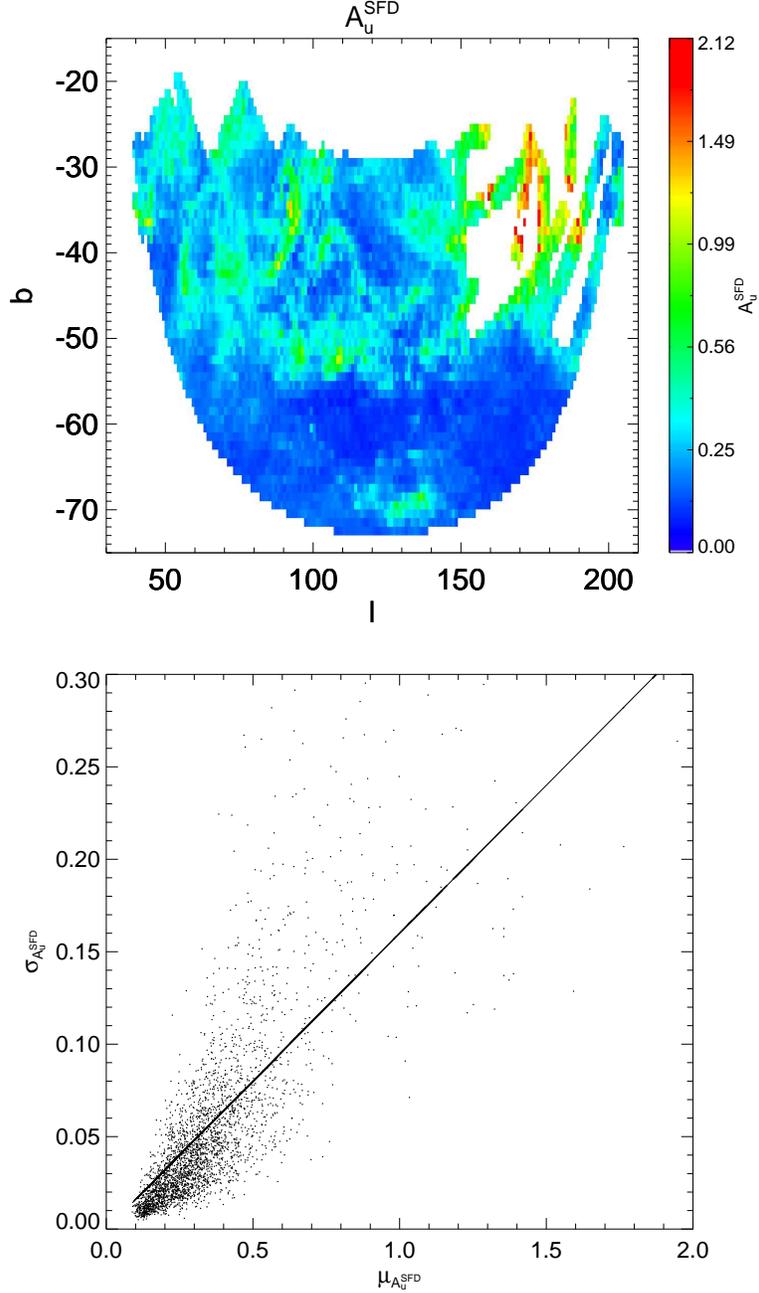}
\caption{The $u$ band SFD extinction ($A_u^{SFD}$) map of 3070 SCUSS fields (top panel). The mean($\mu_{A_u^{SFD}}$) and scatter($\sigma_{A_u^{SFD}}$) of $A_u^{SFD}$ distribution of all SCUSS fields are shown in the bottom panel, where
the solid line shows the relation $\sigma_{A_u^{SFD}}=0.16\mu_{A_u^{SFD}}$.
\label{plotstatic}}
\end{figure}

In this section, we take the SCUSS galaxy catalog and use their number counts to study the Galactic extinction and compare it with the SFD extinction map.
The SFD extinction map in  $u$ band (using ODO extinction curve with $R_V=3.1$) of the 3070 SCUSS/SDSS fields is shown in the top panel of Figure~\ref{plotstatic} (in Galactic coordinate).  As can be seen, because of the high extinction coefficient in $u$ band ($A_u^{SFD}=5.108 E(B-V)^{SFD}$), the expected $A_u^{SFD}$ extinction values span a very wide range even for our high Galactic latitude footprint ($b<-20\arcdeg$), from the lowest $\sim$ 0.11 mag to as high as $\sim$ 1.5 mag for a few regions.

The sky resolution of SFD maps is  $6\arcmin.1$ per pixel. Inside a pixel of the SFD extinction map, the number of galaxies is too small  for statistical study. Even for a $1.08\arcdeg\times1.03\arcdeg$ SCUSS field, the number of galaxies is still not large enough to make high accuracy statistics. Nevertheless, the aim of this study is not to obtain a high resolution Galactic extinction map from galaxy number counts, but to make a systematical comparison of the Galactic extinction with the SFD map. Therefore, we gather as large as the sky coverage with similar Galactic extinction so as to get  high statistical significance.

\begin{figure}
\figurenum{4}
\centering
\includegraphics[scale=0.4]{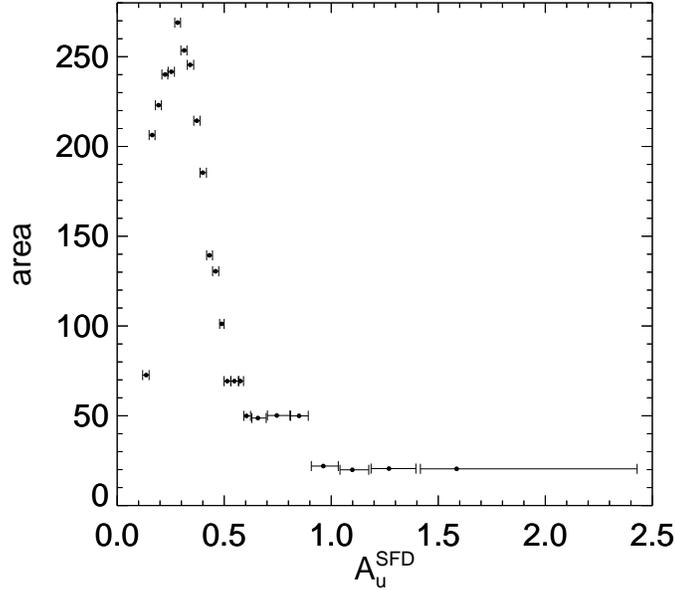}
\caption{The area and mean $A_u^{SFD}$ of the combined regions for galaxy number counts. The error bars show the ranges of the  $A_u^{SFD}$ of each combined region.
\label{subregion}}
\end{figure}

Similar to the approach in \cite{Yasuda2007}, we partition the total SCUSS/SDSS fields into 25 combined  regions according to the  $A_u$ values of the SFD map.
For simplicity, we take each SCUSS field as a unit to build the combined regions if the Galactic extinction inside it is close to a uniform distribution. In specific, we  calculate the mean and standard deviation of  $A_u^{SFD}$ distribution for each SCUSS field (see the bottom panel of Figure \ref{plotstatic}). These  SCUSS fields with normalized standard deviation  smaller than 0.16 (the typical uncertainty of the SFD map) are then considered to have uniform $A_u$ distributions. As can be seen from the  bottom panel of Figure \ref{plotstatic}, the majority of the fields (2134 of 3070) are located below the separation  line  ($\sigma_{A_u^{SFD}}=0.16\mu_{A_u}^{SFD}$), i.e. satisfying the uniform criteria. For the remaining 936 patchy fields, we further divide each of them into $6\times6$ sub-fields ($10.8\arcmin\times10.3\arcmin$). After that, the Galactic extinction distributions inside all these sub-fields satisfy the uniform criteria again. Combining all these uniform fields and sub-fields, we finally get 35830 `units'. We then rank these units according to their mean $A_u$ values.
There are 72 `units' with very low extinction ($A_u^{SFD}<0.12$), which are combined as our reference region (see more details below).
Next, for the `units' with $0.12<A_{u}^{SFD}<0.48$, we set the bin width of $A_u^{SFD}$ being 0.03 mag and get 12 bins. For high extinction regions ($A_u^{SFD}>0.48$), we require each combined region  to have at least 20 square degree sky coverage so as to ensure the statistical significance. With this requirement, we further get 12 combined regions within the range $0.48<A_u^{SFD}<2.43$. In Figure~\ref{subregion}, we show the total areas of all combined regions as function of their $A_u^{SFD}$ ranges.

\begin{figure}
\figurenum{5}
\epsscale{1}
\centering
\includegraphics[scale=0.7]{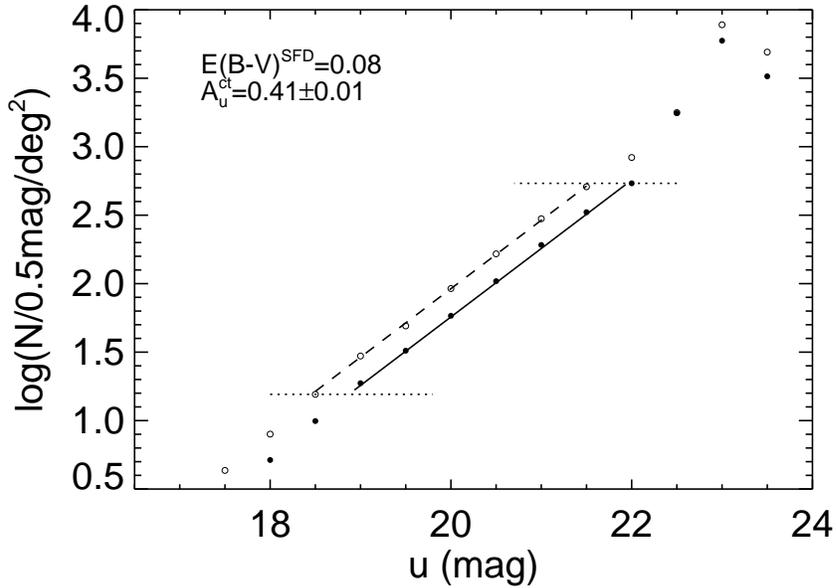}
\caption{The SCUSS $u$-band  galaxy number counts in the reference region (open circles) and the example region ($0.39<A_{u}^{SFD}<0.42$, filled circles). The dashed and  solid lines are the fitting relations of Equation~\ref{eq0} and \ref{eq1} respectively, whose fitting ranges are shown by the dotted lines.
\label{refu}}
\end{figure}

We derive the Galactic extinction in the SCUSS $u$ band by comparing the galaxy number counts of the extincted regions with that of  the reference region \citep{Fukugita2004,Yasuda2007}. For the reference region, we correct the magnitudes of galaxies using the Galactic extinction values from the SFD map.
We show the SCUSS $u$ band galaxy number counts distributions (in terms of the number of galaxies per square degree per 0.5 magnitude)  of an example region and the reference region as the solid and opened circles in Figure~\ref{refu} respectively. The example region is taken from Figure \ref{subregion} where $0.39<A_{u}^{SFD}<0.42$ ($E(B-V)\sim0.08$).

For both the reference and extincted regions, the galaxy number counts in the magnitude range ($18<u<22$) show nice linear relations in the logarithm space.
We fit a linear relation for the reference region with equation
\begin{equation}\label{eq0}
   \log N = \alpha(u-20)+\beta .
 \end{equation}
Then, we fix the $\alpha$ and $\beta$ values and fit a relation
  \begin{equation}\label{eq1}
   \log N = \alpha(u-\Delta M - 20)+\beta .
 \end{equation}
 for the extincted region. Obviously, the fitting parameter $\Delta M$ is the average Galactic extinction of the extincted region.

 It is worth mentioning that, to have self-consistent fitting results, the fitting range of the extincted region should also be shifted from the reference region by an amount of $\Delta M$. More specifically, we set the fitting ranges to be $18.5<u<22.0-\Delta M$ and $18.5+\Delta M<u<22.0$ for the reference and extincted region respectively. The upper magnitude limit $u<22$ mag is chosen where the SCUSS photometric error is smaller than 0.1 mag (Figure~\ref{histerr}), whereas the lower magnitude is selected where the Poisson error of the galaxy number counts is smaller than 5 percent. The fitting result ($\Delta M$) and fitting ranges are finally obtained by iteration. We show the results of the two fitting relations as the dashed line (reference region) and solid line (extincted region) in Figure~\ref{refu} respectively. For this example region, the Galactic extinction $A_u^{ct}$ we obtain is $\Delta M =0.41\pm 0.01$, which is in excellent agreement with the SFD map values ($A_u^{SFD}=0.41$). The error of $\Delta M$ is estimated from the bootstrap sampling of the SCUSS galaxies.

\begin{figure}
\figurenum{6}
\epsscale{1}
\centering
\includegraphics[scale=0.6]{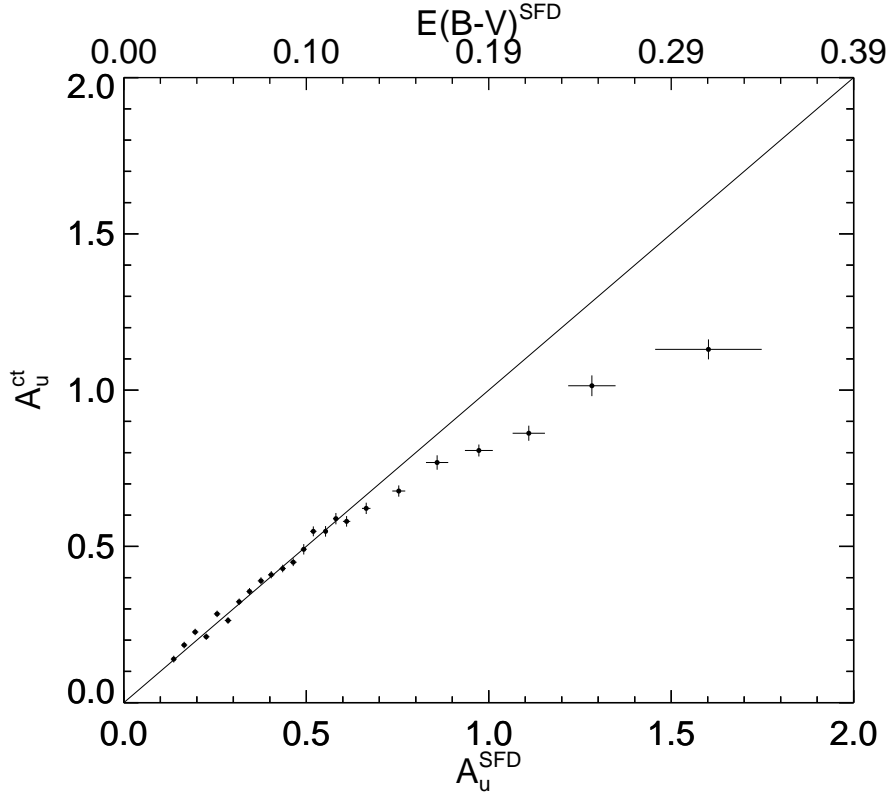}
\caption{The mean and uncertainty of the Galactic extinction $A_u^{ct}$ from galaxy number counts of all 24 regions, which is plotted against $A_u^{SFD}$,  the average extinction values from the SFD map.   The horizontal error bars are the standard deviations of the $A_{u}^{SFD}$ distributions  in each region.
\label{udm}}
\end{figure}

We make galaxy number counts studies for all 24 combined regions and then compare the resulted $A_u^{ct}$ estimations with the SFD values in Figure~\ref{udm}. As can be seen, for low extinction regions ($A_{u}^{SFD} < 0.6 $ mag), the extinction values derived from  galaxy number counts are in excellent agreement with the SFD values. However, for  high extinction regions ($A_u^{SFD}>0.6$ mag), the SFD values are systematically higher.

As we have mentioned,  $A_u^{SFD}$ is derived from  $E(B-V)^{SFD}$, whereas $E(B-V)^{SFD}$ is converted from 100$\mu m$ infrared flux. Both conversions use the standard $R_V=3.1$ ODO extinction curve (see APPENDIX for detail). Therefore, the overestimation of $A_u^{SFD}$ at high extinction regions either comes from the overestimation of  the Galactic dust (conversion from 100$\mu m$ flux)  or a systematical variation of the extinction curve, or both. We leave the discussion of the variation of the extinction curve in Section \ref{sec5}. Assuming the Galactic extinction curve does not show systematical change at high extinction regions, our results indicate that the SFD map have overestimated the Galactic extinction up to 40 percent for $E(B-V)>0.2$ mag regions.

\section{Galactic reddening in \textnormal{$u-r$}} \label{sec4}
In this section, we further use the $u-r$ color distribution of galaxies to test the Galactic reddening of the SFD map. For SCUSS galaxies with $u<23$, except the extreme blue ones ($u-r<0.8$, not used in our statistical study), the $u-r$ color of all other galaxies can be  matched from the SDSS $r$ band photometry catalog (complete to $r<22.2$).

To measure the Galactic reddening of galaxies, a statistical quantity, i.e. the intrinsic color of galaxies need to be well defined. \cite{Strateva2001} studied the $u-r$ colors of SDSS galaxies and found that galaxies can be separated into two populations with separation at  $u-r=2.22$.  In this study, we use the $u-r$ color peak of the blue galaxies ($u-r<2.22$) as a statistical measurement. Unlike the mean or median of the colors of galaxies, the peak of the $u-r$ distribution takes advantage of being unbiased by the incompleteness of the galaxies with extreme colors (e.g. $u-r<0.8$ galaxies in our study). Same as the algorithm in Section \ref{sec3}, we partition the SDSS/SCUSS footprint into 25 regions according to their SFD $E(B-V)$ values . For each region, we select the galaxies with $u<23$ and measure the peak of their $u-r$ distribution. More specifically, we assume that the $u-r$ color of each individual galaxy follows a Gaussian probability distribution function,
\begin{equation}\label{eq2}
P_i(x)=\frac{1}{\sqrt{2\pi\sigma_i}}exp(\frac{-(x-\mu_i)^{2}}{2\sigma_i^{2}})
\end{equation}
where $\mu_i$ and $\sigma_i$ are its observed $u-r$ color and  photometric uncertainty.  The global $u-r$ distribution of the sampling galaxies is then obtained by adding the $u-r$ probability distributions of  individual galaxies
\begin{equation}\label{eq3}
 P(u-r) =\frac{1}{N}\sum\limits_{i=1}^{N}P_{i}(u-r)\,,
\end{equation}
where N is the number of galaxies in consideration. We measure the peak of the resulted $u-r$ distribution with a step of 0.01 mag and estimate its uncertainty with 50 times of bootstrap samplings.

\begin{figure}
\figurenum{7}
\epsscale{1}
\centering
\includegraphics[scale=0.7]{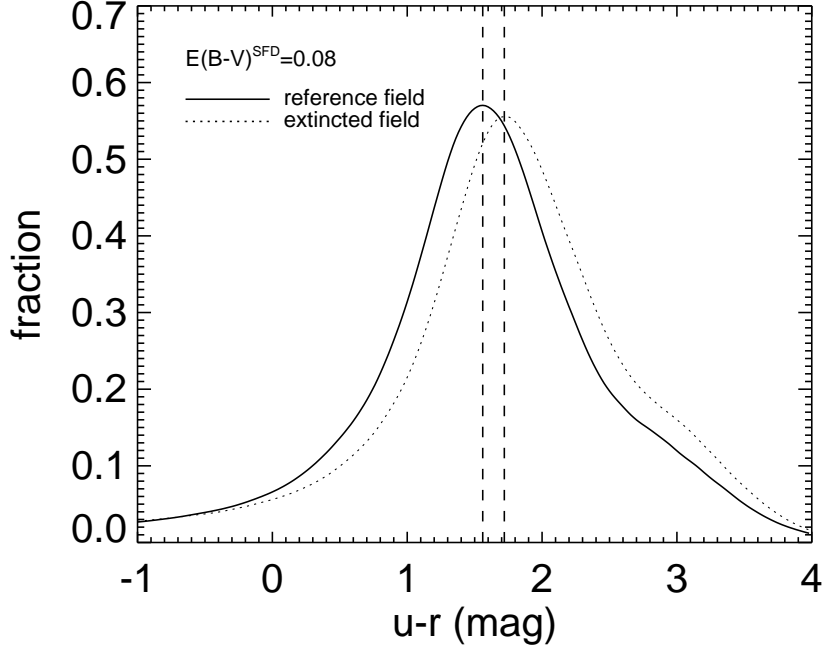}
\caption{The $u-r$ distribution of galaxies. The solid and dotted curves show the results of the reference region and the example region ($E(B-V)^{SFD}=0.08$) respectively. The two dashed lines show the positions of the peaks of two distributions.
\label{ppref}}
\end{figure}

We show the $P(u-r)$ distributions of the galaxies in the reference region and the example region ($0.39<A_{u}^{SFD}<0.42$) with solid and dotted curves in Figure~\ref{ppref} respectively. As expected, the peak of $P(u-r)$  in the example region shows a significant shift to the reference region, which tells us how much the galaxy colors are averagely reddened in the reference region. However, this shift is not the exact average Galactic  reddening value we want to derive.

\begin{figure}
\figurenum{8}
\epsscale{1}
\centering
\includegraphics[scale=0.7]{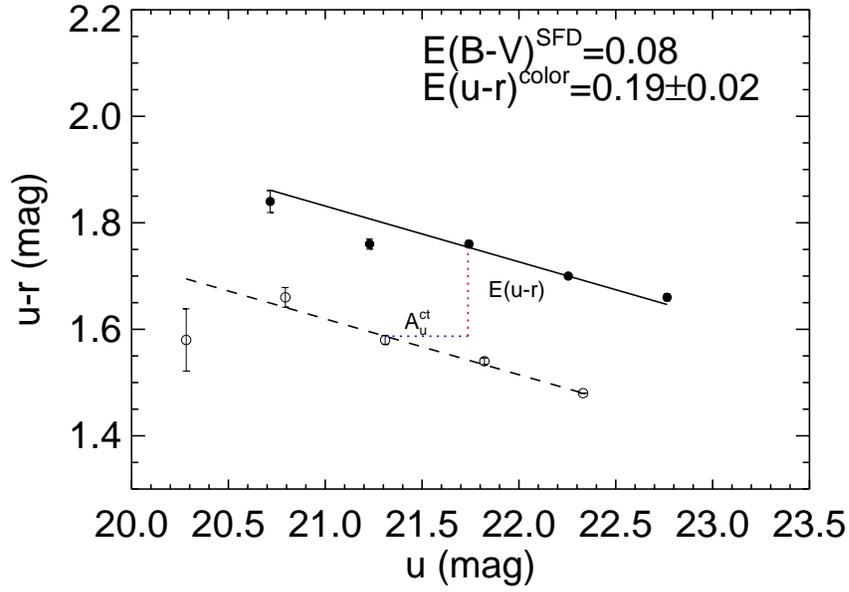}
\caption{The color magnitude relation and Galactic reddening of the example region ($0.39<A_{u}^{SFD}<0.42$). The open and filled circles show the $(u-r)$ -  $u$ relations of the reference region and example region respectively, whereas the dashed  and solid line show the fitting relations of Equation \ref{eq4} and \ref{eq5}.
\label{gdml}}
\end{figure}

There is an intrinsic color-magnitude relation of galaxies, i.e. brighter galaxies also look redder. The Galactic dust not only reddens  galaxy's color but also extincts its flux. Therefore the galaxies with $u<23$ mag in the extincted region are intrinsically brighter than the $u<23$ mag galaxies in the reference region, which makes the galaxies in extincted region intrinsically bluer. To quantify this intrinsic color-magnitude relation, we separate galaxies into different $u$ magnitude bins and measure the peak of their $u-r$ colors accordingly.
As an example, we plot the peak of the $u-r$ distribution as function of $u$ magnitude in Figure~\ref{gdml} for the example and reference regions respectively.
Color-magnitude relations are clearly seen for both samples. We fit a linear relation between the $u-r$ peak color and $u$ magnitude for the reference region

 \begin{equation}\label{eq4}
 u-r  = \alpha^\prime (u -21.5)+\beta^\prime
 \end{equation}

 and then fit the corresponding relation

  \begin{equation}\label{eq5}
 u-r + E(u-r) = \alpha^\prime (u-A_u -21.5)+\beta^\prime
 \end{equation}
for the extincted region, where $\alpha^\prime$ and  $\beta^\prime$ are fixed to the fitting values of Equation 5. In Equation \ref{eq5}, $E(u-r)$ and $A_u$ are the average reddening and extinction of the given extincted region respectively. For each extincted region, the average extinction $A_u$ has already been obtained from the galaxy number counts in Section \ref{sec3} (Figure~\ref{udm}). For consistence, we set the fitting ranges of Equations \ref{eq4} and \ref{eq5} to be $20.0 <u<23-A_u$ mag and $20.0+A_u <u<23$ mag respectively. The fitting ranges of Equation \ref{eq4} and \ref{eq5} are fainter than those of Equation \ref{eq0} and \ref{eq1}, which is due to the fact that we need more number of galaxies to constrain the $u-r$ color peak.  The average Galactic reddening of the example region obtained from Equation \ref{eq5} is $E(u-r)=0.19\pm0.02$, which is also in excellent agreement with the SFD value $E(B-V)=0.08$ for the standard $R_V=3.1$ ODO extinction curve.

\begin{figure}
\figurenum{9}
\epsscale{1}
\centering
\includegraphics[scale=0.6]{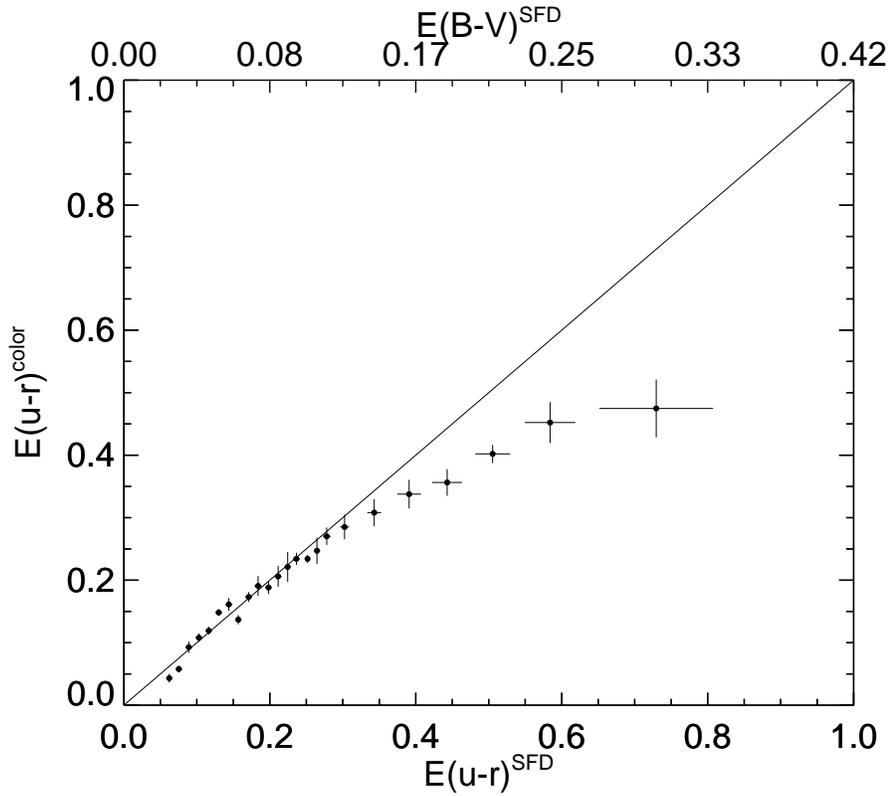}
\caption{The Galactic reddening $E(u-r)^{color}$ from color distributions of galaxies in all  24 regions, which is plotted against the mean $E(u-r)^{SFD}$ from  SFD map.
\label{eur}}
\end{figure}
The average Galactic reddening $E(u-r)$ of all 24 regions is shown  in Figure~\ref{eur}. In this plot, the $E(u-r)$ from  color distribution is plotted against the mean $E(u-r)$ from  SFD map, which is converted from $E(B-V)^{SFD}$ using the standard $R_V=3.1$ ODO extinction curve. We see that the $E(u-r)$ values from galaxy colors are in good consistence with the SFD results at the low reddening regions ($E(B-V)^{SFD}<0.12$ ). Again, for high reddening regions ($E(B-V)^{SFD} > 0.12$), the SFD reddening values  deviate from the galaxy color measurements systematically. Such results are very close to the galaxy number counts results shown in Figure~\ref{udm}.

\begin{figure}
\figurenum{10}
\epsscale{1}
\centering
\includegraphics[scale=0.6]{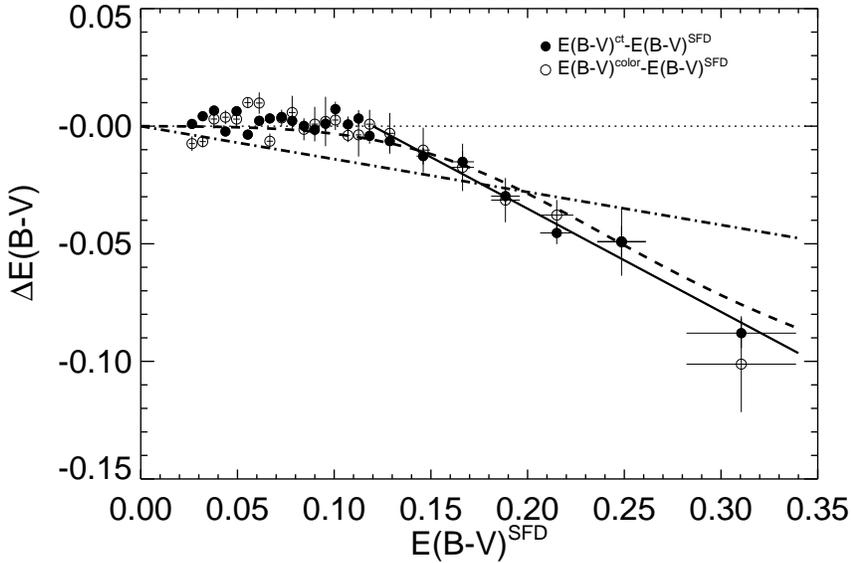}
\caption{The differences of the $E(B-V)$ values between our measurements and SFD map as function of $E(B-V)^{SFD}$. The open and filled circles show the results from $u-r$ galaxy colors and $u$ band galaxy number counts respectively. The dotted, dashed, dotted-dashed and solid lines are the $\triangle E(B-V)=0$, \cite{Yasuda2007} results, $\triangle E(B-V)=-0.14E(B-V)$ of \cite{Schlafly2011} and the linear fitting relation of Equation \ref{eq6}. }
\label{deltebv}
\end{figure}

To have a better combination of the results from Figure~\ref{udm} and Figure~\ref{eur}, we plot the differences of the $E(B-V)$ values between our measurements and the SFD map as function of $E(B-V)^{SFD}$ in Figure~\ref{deltebv}.  The $A_u^{ct}$ and $E(u-r)^{color}$ are both converted back to $E(B-V)$ values using the ODO $R_V=3.1$ extinction curve. The two methods show very consistent results that the SFD map overestimates the reddening values at high extinction regions systematically. Such an overestimation can be approximated by a linear relation
 \begin{equation}
 \Delta E(B-V)= 0.43[E(B-V)^{SFD}-0.12]\label{eq6}
 \end{equation}
when $E(B-V)^{SFD} >0.12$, which is shown by the solid line in Figure~\ref{deltebv}. We also plot the fitting relation of \cite{Yasuda2007} (their Equation 2) and the result of \cite{Schlafly2011} ($E(B-V)=0.86E(B-V)$) as the dashed and dotted-dashed lines respectively for comparison. Our result is in good agreement with \cite{Yasuda2007}.

Both of our study and the study of  \cite{Yasuda2007} use the galaxy number counts whereas the study of \cite{Schlafly2011}  uses the blue tip of the stellar locus to constrain the Galactic reddening. Moreover, the footprint of our study and of \cite{Yasuda2007} are both in south Galactic cap region (although our footprint is much larger), whereas the study of \cite{Schlafly2011} is mainly in the north Galactic sky. Therefore, the different  $\Delta E(B-V)$ correction of \cite{Schlafly2011} may either come from the systematics of the methods or from the different reddening properties in Galactic coordinates.

\section{Discussion: Galactic extinction curve} \label{sec5}

In Section \ref{sec3} and \ref{sec4}, all Galactic extinction and reddening  from SFD map are derived using the standard ODO $R_V=3.1$ extinction curve.
However, as shown detailed in APPENDIX, all these values are model (extinction curve) dependent.
Whether the inconsistent results with SFD map shown at high extinction regions could be reconciled by introducing extinction curves with other $R_V$ values or other model of extinction curve under the assumption that the dust emission provide by SFD is correct?

\subsection{Extinction curve with different $R_V$ values}
\begin{figure}
\figurenum{11}
\epsscale{1}
\centering
\includegraphics[scale=0.5]{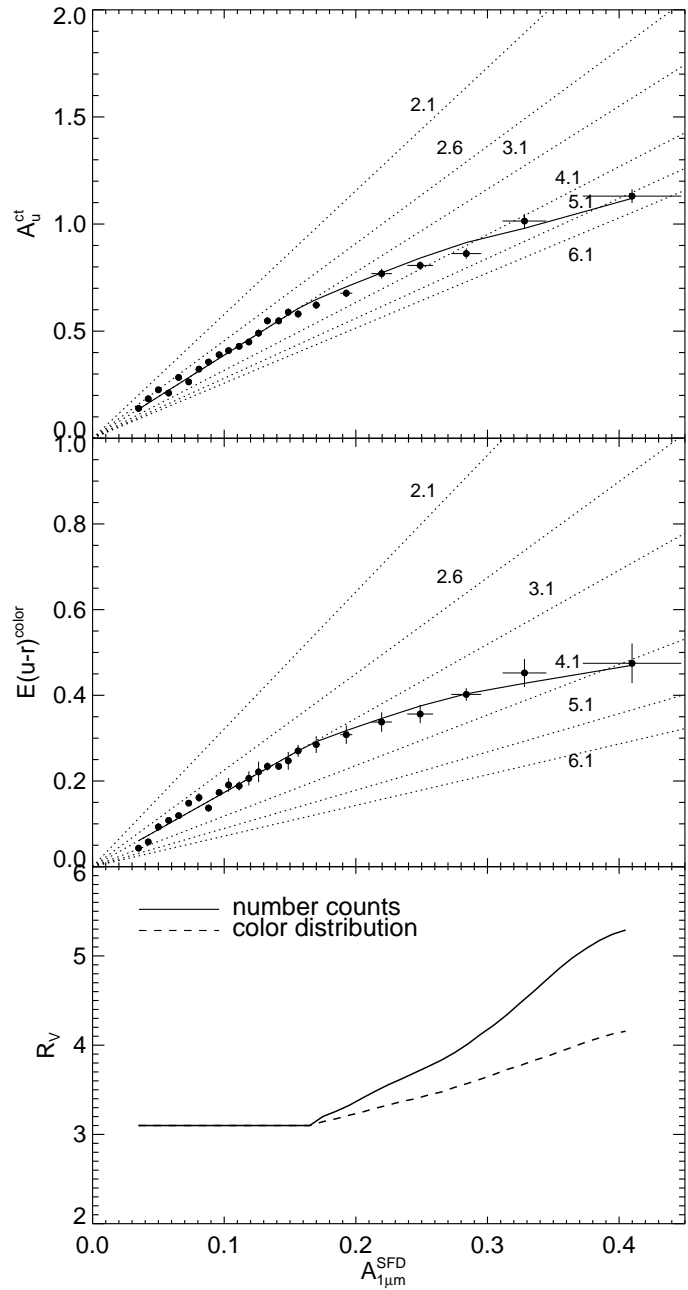}
\caption{The Galactic extinction $A_u^{ct}$ (top), reddening $E(u-r)^{color}$ (middle) and the required $R_V$ values (bottom) as function of $A_{1\mu m}^{SFD}$. The dotted lines in the top two panels show the predicted $A_u$ and $E(u-r)$ values as function of $A_{1\mu m}$ for ODO extinction curves with different $R_V$ values. In the bottom panel, the solid and dashed lines show the required $R_V$ values as function of  $A_{1\mu m}$  when the predicted $A_u$ and  $E(u-r)$  values are forced to be consistent  with the observational results(the solid lines in the top two panels respectively).
\label{fitrv}}
\end{figure}
As we have shown in Figure~\ref{udm} and Figure~\ref{eur}, the Galactic extinction and reddening are systematically overestimated by the SFD map when using standard $R_V=3.1$ extinction curve.
However, this overestimation may also be explained by using an extinction curve with higher $R_V$ values. Indeed,  some studies have shown that $R_V$ increases towards the dense star forming regions \citep[e.g.][]{Savage1979,Cardelli1988,Wang2013}.

To test this idea, we convert the $E(B-V)$ values of  SFD map back to the  Galactic extinction at near-infrared wavelength  1$\mu m$ ($A_{1\mu m}$), which is  believed to be no more dependent on the Galactic extinction curve. We plot $A_u^{ct}$ and $E(u-r)^{color}$ against $A_{1\mu m}$ in the top and middle panel of  Figure~\ref{fitrv} respectively.
At given $A_{1\mu m}$, $A_{u}$ and $E(u-r)$ can be easily predicted for the extinction curves with different $R_V$ values (see APPENDIX for detail).


As expected, at low extinction regions (small $A_{1\mu m}$), our results are in good agreement with the $R_V=3.1$ extinction curve. While at high extinction regions, it seems that the inconsistence between our results and  the SFD values can be reconciled by using the extinction curves with higher $R_V$ values. To further test this idea, we force the $A^{ct}_u$ and $E(u-r)^{color}$ values to be consistent with the $A^{SFD}_{1\mu m}$ values (the two solid lines in the top and middle panel of Figure \ref{fitrv}) and calculate the required $R_V$ values in turn, which are shown as the solid and dashed line in the bottom panel of Fig. 11 respectively.  As can be seen, we can not get  self-consistent results between the galaxy number counts and color distribution by only varying the $R_V$ values of the extinction curves.

\subsection{ODO VS FM extinction curve}

\begin{figure}
\figurenum{12}
\epsscale{1}
\centering
\includegraphics[scale=0.6]{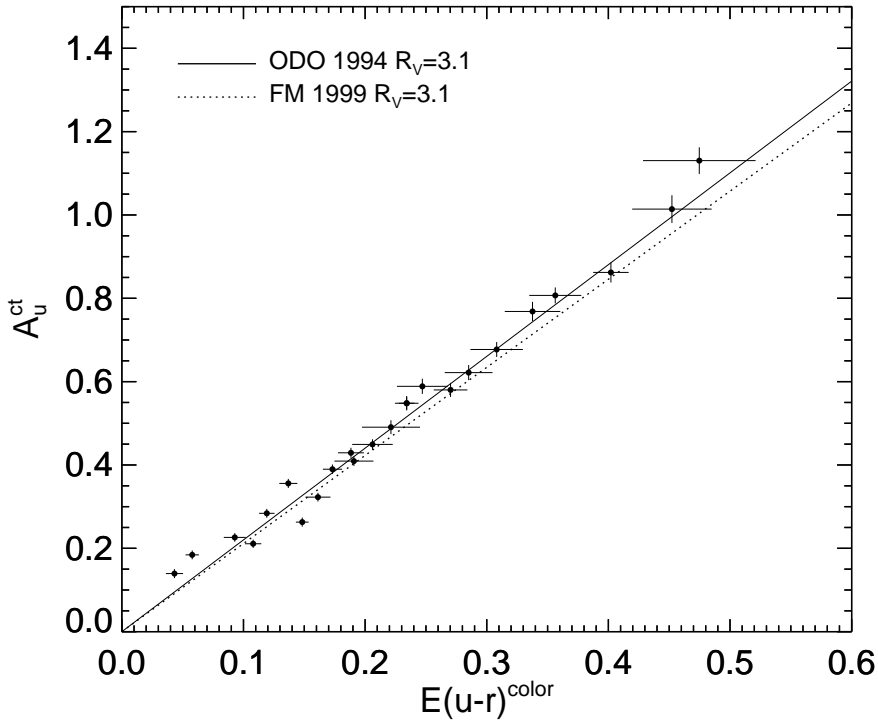}
\caption{The correlation between the Galactic extinction $A_u^{ct}$ from galaxy number counts and reddening  $E(u-r)^{color}$ from color distribution. The solid and  dotted lines show the relations between $A_u$ and $E(u-r)$ from  $R_V=3.1$ ODO and FM extinction curves respectively.
\label{cteur}}
\end{figure}

The $A_u^{ct}$ and $E(u-r)^{color}$ values, besides  both being model (extinction curve) independent measurements, also provide a strong constraint on the shape of the Galactic extinction curve when combined together. In Figure \ref{cteur}, we plot $A_u^{ct}$ against $E(u-r)^{color}$ values and compare their relation with the  results of the two most frequently used  extinction curves, the ODO and \citet{1999Fitzpatrick}(FM) one. For the $R_V=3.1$ ODO and FM extinction curves, the slopes of the $A_u$ and $E(u-r)$ relation, i.e. the $A_u/E(u-r)$ values, are 2.203 and 2.114 respectively. As can be seen, the $A_u^{ct}$ and $E(u-r)^{color}$ relation is in excellent agreement with  the  ODO extinction curve, which is also significantly better than the  FM one.

In order to further quantify the differences between the ODO and FM extinction curves, we make a linear fitting on the relation of  $A_u^{ct}$ and $E(u-r)^{color}$. Considering the uncertainty of the zero reddening of the reference region, we do not set the intercept of the fitting relation to be zero but set it as a free parameter. The fitting results we get are $A_u=(2.194\pm 0.031)E(u-r) + (0.018\pm0.006)$. The intercept value from fitting is in agreement with zero, while the slope value confirms that the ODO extinction curve is better than the FM one in about $2.5\sigma$ level.

The excellent agreement with the ODO extinction curve of the $A_u$ and $E(u-r)$ relation strongly  implies that the overestimation of the Galactic reddening of the SFD values at high extinction regions is indeed caused by the over-estimation of the infrared flux (or its conversion factor) and  unlikely be explained by the variation of the extinction curve.

\section{Summary and Conclusion} \label{sec6}
In this paper, we use the galaxy catalog of South Galactic Cap $u$-band Survey to study the integral Galactic extinction and reddening using galaxy number counts and color distribution respectively. Benefited from the sensitivity of $u$ band to dust extinction and SCUSS depth (1.5 mag deeper  than SDSS), we have obtained the average Galactic extinction $A_u^{ct}$ and reddening $E(u-r)^{color}$ in the south Galactic cap region with unprecedented statistical accuracy. By combining $A_u^{ct}$ and $E(u-r)^{color}$ results together, we constrain the shape of Galactic extinction curve and find that it is in excellent agreement with the standard $R_V=3.1$ ODO extinction law. Our results also confirm that the SFD map overestimates the Galactic extinction at higher extinction regions ($E(B-V)>0.12$). This overestimation is up to the level of about 40 percent and can be corrected using a linear relation $\Delta E(B-V)= 0.43[E(B-V)^{SFD}-0.12$]. As already been discussed in \cite{{Yasuda2007}}, this overestimation may be caused by the underestimation of the dust temperature of the 100$\mu m$ emission when constructing the SFD extinction map.

 The footprint in this study located in  high Galactic region where the Galactic reddening  does not reach very high extinction value as that in the Galaxy disk. The maximum $E(B-V)$ value we can statistically probe is about 0.35. Therefore, the overestimation and correction of the SFD map we find at $E(B-V)>0.12$ (Equation \ref{eq6}) may only be valid in the range $0.12<E(B-V)<0.35$. Moreover, although our study covers a large footprint of the south Galactic cap region ($\sim$ 3000 square degrees), we caution that this conclusion may also only be valid in the studied areas. As shown by \cite{Welty1992} and \cite{1999Fitzpatrick}, the Galactic extinction pattern (dust properties) may vary significantly in difference positions. Independent measurements of the Galactic extinction and reddening (as in our study) in very highly extincted regions and with all sky coverage worth further studies.

\acknowledgments
LL thanks Hengxiao Guo and Shuai Feng for help of useful discussions. This work was supported by the National Natural Science Foundation of China (NSFC) with the Project Numbers  11573050,  11433003,  the ``973 Program'' 2014 CB845705 and the Strategic Priority Research Program ``The Emergence of Cosmological Structures'' of the Chinese Academy of Sciences (CAS; grant XDB09030200). .YFT is supported by NSFC with No.11303070.

\appendix\label{ap}

\section{Galactic extinction and reddening from SFD map}

In SFD, the $100\mu$m thermal emission is converted to reddening $E(B-V)$ by $E(B-V)=pD^T$, where $p$ is a calibration coefficient, and $D^T$ represents the point source-subtracted $IRAS$-resolution 100 $\mu$m emission corrected by a temperature factor. In the released SFD map of $E(B-V)$, $p$ is calculated using the standard $R_V=3.1$ ODO extinction law and then calibrated using the colors of elliptical galaxies.

For the extinction curves other than the standard ODO $R_V=3.1$ one, the released SFD $E(B-V)$ map can not be directly used to calculate the Galactic extinction and reddening values.
Alternatively,  the $E(B-V)$ values of the SFD map can be converted back to the Galactic extinction in near-infrared wavelength (e.g. 1$\mu m$ ), which is believed to be dependent on the extinction curve very weakly. Then, the Galactic extinction curves can be  expressed as
  \begin{equation}
k(\lambda)=A(\lambda)/A(1\mu m) \,.
  \end{equation}
For a given $A_{1\mu m}$ value, the Galactic extinction  in any given band $b$ is calculated by
  \begin{equation}\label{A1}
  A_b=-2.5log[\frac{\int d\lambda W_b(\lambda)S(\lambda)10^{-k(\lambda) A_{1\mu m}/2.5}}{\int d\lambda W_b(\lambda)S(\lambda)}]
 \end{equation}
 \begin{figure}
\figurenum{13}
\epsscale{1}
\centering
\includegraphics[scale=0.7]{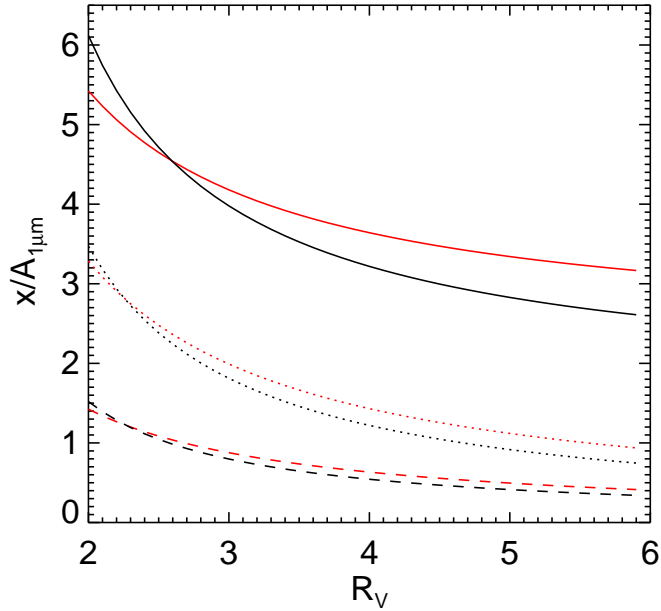}
\caption{Extinction coefficient normalized to $A_{1\mu m}$ as function of $R_V$. The black solid, dotted and dashed lines show $A_u/A_{1\mu m}$, $E(u-r)/A_{1\mu m}$, and $E(B-V)/A_{1\mu m}$ using the ODO extinction law, while the results from the FM extinction law are shown by the red ones.
\label{klam}}
\end{figure}
where $W(\lambda)$ is the transmission curve of  the given band, and $S(\lambda)$ is the source spectrum.
The transmission curve is  a convolution of the filter transmission, CCD quantum efficiency, and  atmospheric extinction, which are publicly available for all photometric bands\footnote{The SCUSS $u$ band transmissions is available at\\\url{http://batc.bao.ac.cn/BASS/doku.php?id=scuss:facilities:homefilter}.\\ The SDSS $r$ band data is avaliable at\\{\url{https://www.sdss3.org/instruments/camera.php
}} }. For $S(\lambda)$, we take the mean galaxy spectrum from SDSS\url{}. We have tested that a more realistic source spectrum make negligible changes to all of our results.

 For a given extinction curve, we calculate the extinction coefficient  $k(b)=A_b/A_{1\mu m}$ for all the related bands.  The resulted  $A_u/A_{1\mu m}$, $E(u-r)/A_{1\mu m}$, and $E(B-V)/A_{1\mu m}$ values are plotted as function of $R_V$ values for FM and ODO extinction curves in Figure 13.

\bibliographystyle{aasjournal}
\bibliography{j1923_kennedy.bib}

\end{document}